\documentclass[journal=ancac3,manuscript=article]{achemso}

\usepackage[version=3]{mhchem} % Formula subscripts using \ce{}

\author{Shijun Zhao}
\email{zhaosj@pku.edu.cn;}
\affiliation[Center for Applied Physics and Technology]
{HEDPS, Center for Applied Physics and Technology, Peking University, Beijing 100871, P. R. China}
\alsoaffiliation[College of Engineering]
{College of Engineering, Peking University, Beijing 100871, China}
\author{Wei Kang}
\affiliation[Center for Applied Physics and Technology]
{HEDPS, Center for Applied Physics and Technology, Peking University, Beijing 100871, P. R. China}
\alsoaffiliation[College of Engineering]
{College of Engineering, Peking University, Beijing 100871, China}

\title{The potential applications of phosphorene as anode materials in Li-ion batteries}

\begin{document}

\pagebreak

%\begin{abstract}

\textbf{Abstract} The capacity and stability of constituent electrodes determine the performance of Li-ion batteries. In this study, density functional theory is employed to explore the potential application of recently synthesized two dimensional phosphorene as electrode materials. Our results show that Li atoms can bind strongly with phosphorene monolayer and double layer with significant electron transfer. Besides, the structure of phosphorene is not much influenced by lithiation and the volume change is only 0.2\%. A semiconducting to metallic transition is observed after lithiation. The diffusion barrier is calculated to 0.76 and 0.72 eV on monolayer and double layer phosphorene. The theoretical specific capacity of phosphorene monolayer is 432.79 mAh/g, which is larger than other commercial anodes materials. Our findings show that the high capacity, low open circuit voltage, small volume change and electrical conductivity of phosphorene make it a good candidate as electrode material.

%\end{abstract}


\newpage
%%%%%%%%%%%%%%%%%%%%%%%%%%%%%%%%%%%%%%%%%%%%%%%%%%%%%%%%%%%
\section{INTRODUCTION}

Due to the high reversible capacity, high energy density and good cycle life, lithium-ion batteries (LIBs) have become predominant battery technology in recent years.The performance of LIBs relies greatly on the Li capacity and cycle rate of anode materials. Owing to the large surface-to-volume ratio and unique electronic properties distinguished from their bulk counterparts, two-dimensional (2D) materials have attracted a great deal of attentions in the filed of LIBs.\cite{Novoselov2004,Geim2007} Actually, the possibility of various 2D sheets used as anodes in LIBs has been explored from the first 2D graphene sheet to recetly synthesized transion-metal-dichalcogenides (TMD) and MXenes.\cite{Pollak2010,Tritsaris2013}\cite{Tang2012} In view of their flat structures, a fast Li diffusion and large Li capacity can be achieved. However, the weak binding of Li with graphene and other 2D sheet hinders its further applications in LIBs.\cite{Pollak2010}

Very recently, a new 2D semiconducting material with a direct bandgap called phosphorene has been successfully fabricated.\cite{Li2014,Liu2014,Reich2014} As in the case of other 2D sheets, a high mobility is found in phosphorene-based field effect transistors. Besides, the electronic properties of phosphorene are demonstrated to be dependent on its thickness. The outstanding properties of phosphorene have aroused considerable
interest and much research efforts have been devoted to explore its various applications.\cite{Dai2014}

%The structure of phosphorene is consisting of P atoms bonded with three adjacent P atoms, forming a puckered honeycomb structure.

In this work, we employ first-principles calculations based on density functional theory (DFT) to study the adsorption and diffusion of Li on phosphorene to assess the suitability of phosphorene as a host material for LIBs. The free-standing single-layer and double-layer structures of phosphorene are considered. Our calculations indicate that Li atoms bind stronger on double layer phosphorene than monolayer. The lowest diffusion barrier of Li atoms is calculated as 0.76 and 0.72 eV on monolayer and double layer phosphorene. The theoretical capacity of Li is found to be 432.79 mAh/g and 324.59 mAh/g for monolayer and double layer phosphorene, which is larger than other commercial anodes used in LIBs. These results suggest phosphorene holds great potentials in LIBs.

\section{Computational methods}

All calculations have been carried out based on DFT as implemented in the QUANTUM-ESPRESSO package.\cite{Giannozzi2009} The exchange correlation energy is described by the generalized gradient approximation (GGA) in the scheme proposed by Perdew-Burke-Ernzerhof (PBE).\cite{Perdew1996} Electron-ion interactions are described by projector augmented wave pseudopotentials.\cite{Blochl1994} A plane-wave basis set with a cutoff of 50 Ry is used. In order to seek for possible magnetic ground states, a random initial magnetization is used for spin-polarized calculations. For the calculations of phophosene double layer, the van der Waals interaction is included using a dispersion correction term with DFT-D method.\cite{Grimme2006}

The phosphorene is modeled by 2$\times$2 supercell to investigate the properties of Li adsorption. Periodic boundary condition (PBC) is applied, in which the size in the perpendicular direction is large enough (larger than 10~\AA) to avoid spurious interactions induced by periodic images. The positions of atoms are fully optimized using Broyden-Fletcher-
Goldfarb-Shanno (BFGS) method with an energy convergence of 10$^{-6}$ Ry between two consecutive self-consistent steps and a force convergence of 4$\times$10$^{-4}$ Ry/Bohr. The Brillouin zone is sampled using a 16$\times$12$\times$1 Monkhorst-Pack grid. After structural optimization, the density of states (DOS) and electronic properties of the layers are calculated
using denser 20$\times$16$\times$1 k points.

The binding energy per Li for the adsorption of $n$ Li atoms is defined as
\begin{equation}
E_b=(E_{\mathrm{phosphorene}}+nE_{\mathrm{Li}}-E_{\mathrm{phosphorene}+Li_n})/n,
\end{equation}
where $E_{\mathrm{phosphorene}}$ is the total energy of the pristine phosphorene layer, $E_{\mathrm{phosphorene+Li}_n}$ is the total energy of phosphorene layer adsorbed with $n$ Li atoms and the $E_{\mathrm{Li}}$ is the energy of an isolated Li atom. Under this definition, a higher value of $E_b$ means stronger binding of Li to phosphorene layer. To denote the concentration of Li on phosphorene, we use the notation of Li$_x$P$_{1-x}$ as previous studies.\cite{Osborn2012a,Tritsaris2013a} Note that the binding energy defined here can also be interpreted as the open circuit voltage (OCV) used in battery fields if the entropy effects are neglected.\cite{Er2014}

%In order to examine the stability of lithiated Ti$_2$C monolayer and double layer, a series of molecular dynamic simulations are also performed. The temperature is maintained at room temperature of 300 K with a time step of 1 fs. The total duration of the simulation is typically 3 ps.

\section{Results and discussion}
\subsection{Monolayer phosphorene}

In order to validate our approaches, we first calculate the lattice constant of phosphorene by optimizing the in-plane cell dimensions and all atomic positions. The obtained lattice constants are 4.623 and 3.298~\AA, which are in line with previous results.

\begin{figure}[!htb]
   \begin{center}
   \includegraphics[width=0.4\textwidth]{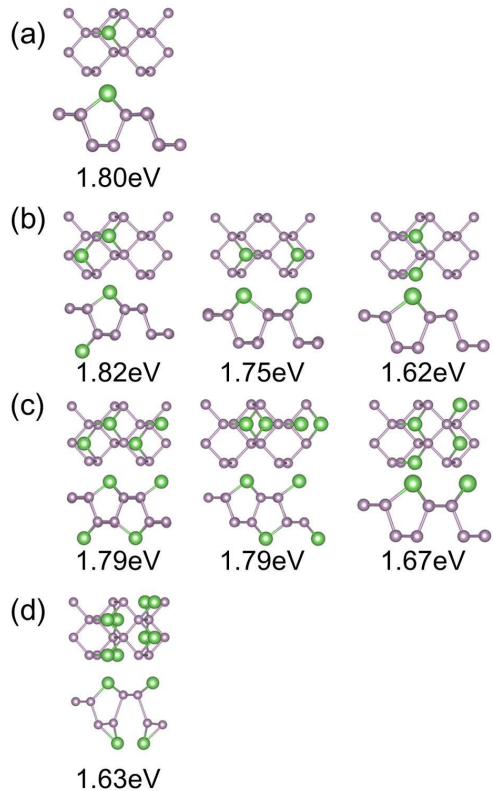}
  \end{center}
   \caption{Optimized stable configurations of $n$ Li atoms adsorption on phosphorene monolayer: (a) $n$=1, (a) $n$=2, (a) $n$=4, (a) $n$=8.}
   \label{mono}
\end{figure}

The capacity of Li is vital for the application of LIBs. We then investigate the adsorption of Li with different concentrations on phosphorene monolayer. The optimized stable adsorption configurations and corresponding binding energies of Li on phosphorene are provided in Fig.~\ref{mono}. The number of Li atoms is 1, 2, 4 and 8, which means that the x value in Li$_x$P$_{1-x}$ is 0.06, 0.11, 0.2 and 0.33, respectively.

%++++++++++++++++++++++++++++++++++++++++++++++++++++++++++++++++++++++
%++++++++++++++++++++++++++++++++++++++++++++++++++++++++++++++++++++++

The most stable adsorption site for a single Li atom is just above the center of the triangle consisting of three P atoms in the surface, which results in a binding energy of 1.80 eV. The bonding distance of Li with the three nearest P is 2.54, 2.54 and 2.45~\AA, respectively. For two Li atoms, the most stable configuration contains one Li atom above the phosphorene layer and the other Li below, both  above the triangle site. The adsorption of two Li atoms both at one side is less stable. Likewise, four Li atoms tend to resides both sides of phosphorene. The most stable adsorption configuration for eight Li atoms occurs when they situated above or below all the available triangle centers. Adding additional Li atoms will push the adsorbed Li into another plane and lead to small binding energies due to increased repulsion between Li atoms in the Li adlayer. Therefore, we take the adsorption of eight Li atoms as fully lithiated phosphorene, corresponding to Li$_{0.33}$P$_{0.67}$.

It is demonstrated in Fig.~\ref{mono} that the binding energy of Li varies in a narrow range less than 0.2 eV with the increase of concentrations. The small variation is also found in other 2D materials.\cite{Tritsaris2013b,Osborn2012a} The high binding energy combining with small bonding distance between Li and P indicates that Li is chemisorbed on phosphorene. Nevertheless, the adsorption of Li atoms leads to little geometric changes to phosphorene which suggesting that the structure of phosphorene is robust against Li insertion.

\begin{figure}[!htb]
   \begin{center}
   \includegraphics[width=0.5\textwidth]{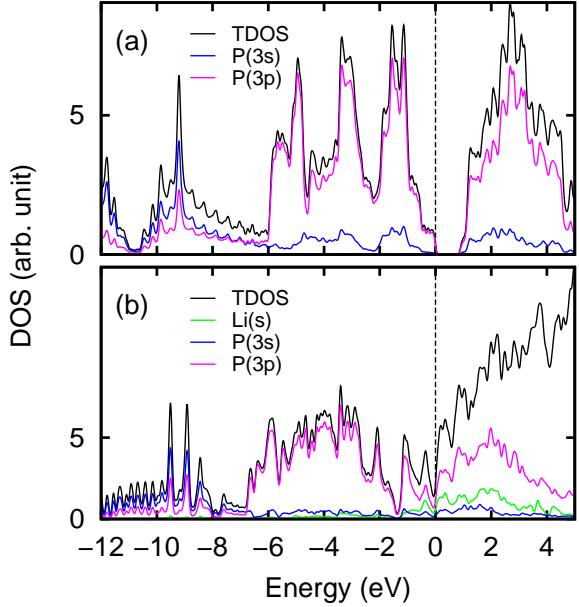}
  \end{center}
   \caption{Total density of states and projected density of states of pristine phosphorene (a) and fully lithiated phosphorene monolayer (b).}
   \label{mono-dos}
\end{figure}

To get further insight into the interactions between Li and phosphorene monolayer, we have made an analysis of the density of states and present the results in Fig.~\ref{mono-dos}. It is shown that the pristine phosphorene is semiconducting with a band gap of 1 eV, in line with previous calculations. The conduction bands are mainly composed of P(3$p$) states. After fully lithiation corresponding to the configuration shown in Fig.~\ref{mono}(d), a metallic transition is observed as there are substantial electron states occupied at the Fermi level. This indicates that the adsorbed Li atoms denote electrons to the phosphorene system. Based on charge analysis, the transferred electron is 0.62 e$^-$ from Li atoms above and 0.51 e$^-$  from Li atoms below. For the adsorption of a single Li atom as shown in Fig.~\ref{mono}(a), the electron transfer is 0.71 e$^-$. The large electron transfer indicates that Li atoms are strongly polarized after the adsorption.

%++++++++++++++++++++++++++++++++++++++++++++++++++++++++++++++++++++++
%++++++++++++++++++++++++++++++++++++++++++++++++++++++++++++++++++++++

\begin{figure}[!htb]
   \begin{center}
   \includegraphics[width=0.5\textwidth]{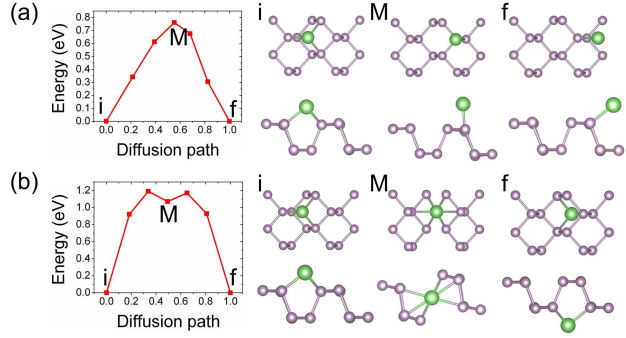}
  \end{center}
   \caption{Diffusion barrier and path of Li atoms on phosphorene monolayer.}
   \label{mononeb}
\end{figure}

Besides the adsorption properties, facile motion of Li is essential for the performance of anode material in LIBs. We then investigate the diffusion path and energy barrier of Li on phosphorene monolayer and the results are presented in Fig.~\ref{mononeb}. It has been shown that he stable adsorption site for Li is just above the center of triangle. As Li moves between two such adsorption sites, it must pass over a P atom in the surface which results in an energy barrier of 0.76 eV. Another Li diffusion path is from the adsorption site above phosphorene to the other side, in which the Li atom passes through the phosphorene layer. The energy barrier is 1.19eV with a local energy-minimum point in the middle of the path. In this intermediate configuration, Li atom is situated at the center of eight P atoms. Since this energy barrier is larger compared to in-plane diffusion, the diffusion of Li is preferable on the surface of phosphorene monolayer.

%++++++++++++++++++++++++++++++++++++++++++++++++++++++++++++++++++++++
%++++++++++++++++++++++++++++++++++++++++++++++++++++++++++++++++++++++

\subsection{Double phosphorene}

We now turn to the properties of Li adsorption and diffusion on double layer phosphorene. There are three possible stacking orders for double layer phosphorene and previous calculations have demonstrated that the AB-stacking is the most energy favorable configuration. Therefore, we chose the this structure to represent phosphorene double layer to investigate its interaction with Li atoms. Our calculated lattice constant of AB stacking phosphorene double layer is 4.529 and 3.307 ~\AA, interlayer distance of 3.20~\AA, which is in line with previous reports.\cite{Dai2014}

\begin{figure}[!htb]
   \begin{center}
   \includegraphics[width=0.4\textwidth]{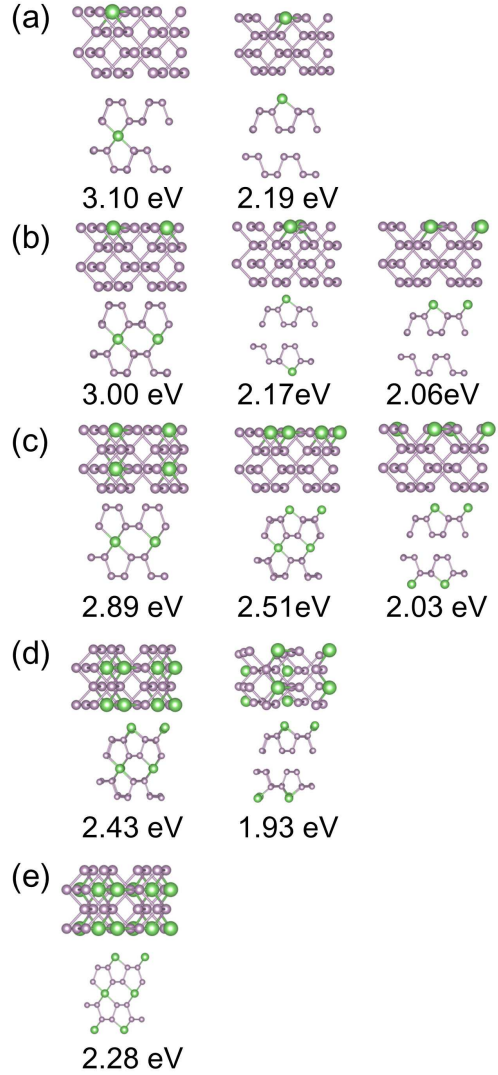}
  \end{center}
   \caption{Optimized stable configurations of $n$ Li atoms adsorption on phosphorene double layer: (a) $n$=1, (a) $n$=2, (a) $n$=4, (a) $n$=8.   }
   \label{double}
\end{figure}

The optimized adsorption configurations and binding energies of Li on double layer phosphorene is provided in Fig.~\ref{double}. The number of Li atoms is
1, 2, 4, 8 and 12, corresponding to $x$ value of 0.03, 0.06, 0.11, 0.20 and 0.27. It is shown that the a single Li atom prefers to reside in the interlayer space, which gives rise to a large binding energy of 3.10 eV. On the other hand, the Li adatom adsorbed on the surface results in lower binding energies. This fact reflects that the interlayer region peculiar to double layer phosphorene provide an unique space for the accommodation of Li atoms, which is beneficial for the improvement of the capacity for LIBs. Likewise, the stable adsorption structure for two Li atoms is that both Li atoms reside in the interlayer region. The adsorption of Li on either surface of double layer phosphorene results in around 1eV lower binding energies. The same rule holds for the adsorption of four Li atoms. With increasing Li contents, the Li atoms firstly occupy the interior of phosphorene double layer and then cover the two external surfaces. In the fully lithiated structure, twelve Li atoms are adsorbed. Further increased Li number leads to the delamination of Li adlayer and wakens the binding energy of Li significantly.

%++++++++++++++++++++++++++++++++++++++++++++++++++++++++++++++++++++++
%++++++++++++++++++++++++++++++++++++++++++++++++++++++++++++++++++++++

By inspecting the binding energies of Li, one noticeable feature is that the Li atoms binding stronger with phosphorene double layer than monolayer owing to their higher binding energies. For example, the binding energy of a single Li atom in double layer is 3.10 eV compared to 1.80 eV in single layer.
Our charge analysis reveals that 0.58 e$^-$ is transferred from Li when it is adsorbed inside phosphorene double layer. For fully lithiated double layer, the charge transfer is 0.56 e$^-$. Nevertheless, the binding energies decrease quickly along with the increase of Li concentrations. The structure of phosphorene is affected little after Li adsorption, indicating that phosphorene is rather stable during lithiated and delithiated process.

\begin{figure}[!htb]
   \begin{center}
   \includegraphics[width=0.5\textwidth]{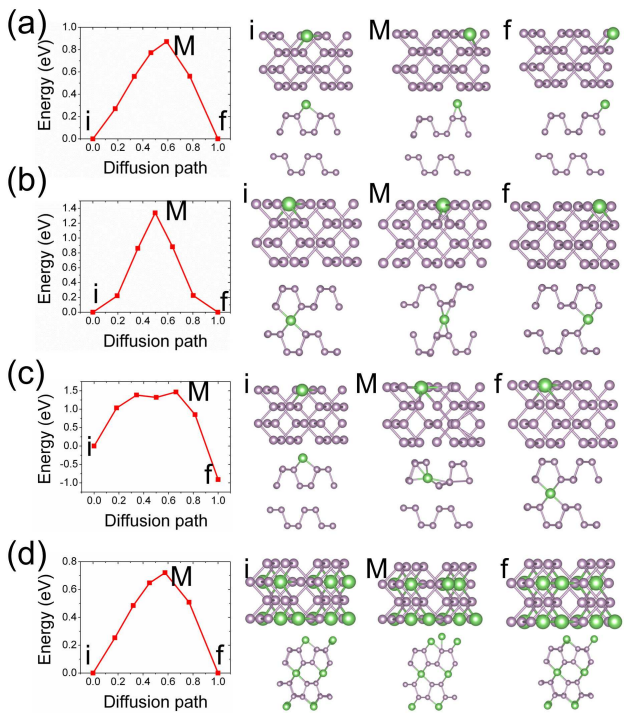}
  \end{center}
   \caption{Diffusion barrier and path of Li atoms on phosphorene monolayer.}
   \label{double-neb}
\end{figure}

The diffusion of Li atoms with low and high concentrations on phosphorene double layer is then investigated and the results are provided in Fig.~\ref{double-neb}. Three diffusion paths are considered for a single Li atom in low-coverage limit. The diffusion of Li above phosphorene double layer has an energy barrier of 0.87 eV by passing over a P atom. As shown in Fig.~\ref{double}, the binding energy of Li inside phosphorene double layer is much larger than that on the surface. When Li atom moves inside the phosphorene double layer, the diffusion barrier increases to 1.34 eV due to steric hindrance. As Li moves from these two adsorption sites, it passes through the phosphorene single layer by overcoming an high energy barrier of 1.47 eV. In this case, the Li atom should break through the P-P bonds and require much energy to pass through the layer. Note that the energy of final state is smaller than the initial state as the interlayer site is more energy favorable. The results thus suggest that in-plane diffusion is easier for Li as observed in phosphorene monolayer. The diffusion of Li at high concentrations is modeled by removal a Li from the fully lithiated phosphorene double layer and then the diffusion barrier is calculated. As shown in Fig.~\ref{double-neb}(d), the barrier for Li diffusion is 0.72 eV as Li passes over P atoms. Compared to the results for phosphorene monolayer, we see that the diffusion properties are not much influence by the increase of layer number.

%++++++++++++++++++++++++++++++++++++++++++++++++++++++++++++++++++++++
%++++++++++++++++++++++++++++++++++++++++++++++++++++++++++++++++++++++

\subsection{Theoretical capacity}

As described above, the fully lithiated phosphorene with stable configuration corresponds to the formulae PLi$_{0.5}$ and PLi$_{0.375}$ for monolayer and double layer respectively. Thus, the theoretical specific capacity provided by the above structure is calculated to be 432.79 and 324.59 mAh/g. This suggests that monolayer phosphorene has larger capacity for Li than double layer. According to our calculated binding energies, it is demonstrated that Li atoms bind stronger on double layer phosphorene compared to monolayer, which indicates that the monolayer phosphorene has lower open circuit voltage and at the same time larger capacity. Thus the phosphorene monolayer is well suitable for the application as anode materials in LIBs.

In comparison to other commercial anode materials, the capacity of phosphorene monolayer is larger than well-studied graphite (372 mAh/g), MXenes such as Ti$_3$C$_2$ (320 mAh/g) and TiO$_2$ polymorphs (200 mAh/g). Therefore, our results demonstrate that phosphorene can be used as anodes in LIBs with low open circuit voltage, high capacity. Besides, the lithaliated phosphorene is metallic which can provide considerable electronic conductivity.

Another important factor to take into account in LIBs is that the possible volume change caused by lithiation and delithiation process. To evaluate this effect, we have optimized the fully lithiated phosphorene monolayer. The results show that the volume change is only 0.2\%, which means that phosphorene is robust against Li insertion and does not suffer from structural changes, which is indispensable in practical applications.

\section{Conclusion}

First-principles calculations based density functional theory have been carried out to investigate the adsorption and diffusion of Li atoms on phosphorene monolayer and double layer to explore its potential applications in LIBs. Our results show that the phosphorene undergoes a transition from semiconducting to metallic upon fully lithiation, which provides additional electric conductivity. The binding of Li atoms is stronger on phosphorene double layer than monolayer. We further show that the lowest diffusion barrier is about 0.76 and 0.87 eV for a single Li atom on phosphorene monolayer and double layer. The theoretical capacity of monolayer is calculated to be 432.79 mAh/g, which is larger than other commercial anodes used in LIBs. Our findings show that the high capacity, low open circuit voltage, small volume change and electrical conductivity of phosphorene make it a good candidate as electrode material.

\begin{acknowledgement}
This work is financially supported by NSAF (Grant No. U1230111), NSFC (Grant No. 91226202 and No. 11274019) and the China Postdoctoral Science Foundation (Grant No. 2014M550561).
\end{acknowledgement}

%\begin{suppinfo}
%\subsection{Supporting Information Available}
%Strain-stress relation of Ti$_2$C. Spin density plot of Li adsorption at strained Ti$_2$C. Adsorption configurations of Li on Ti$_2$C. Configurations of fully lithiated Ti$_2$C layer before and after MD simulations. This information is available free of charge via the Internet at http://pubs.acs.org.
%\end{suppinfo}

%\bibliographystyle{biochem}
\bibliography{pho}

%\pagebreak

%\section{Graphical Table of Contents}

%\begin{figure}[h!]
%	\centering
%	\includegraphics[width=6.0in]{toc.eps}
%\end{figure}

%%%%%%%%%%%%%%%%%%%%%%%%%%%%%%%%%%%%%%%%%%%%%%%%%%%%%%%%%%%
%%%%%%%%%%%%%%%%%%%%%%%%%%%%%%%%%%%%%%%%%%%%%%%%%%%%%%%%%%%
\end{document}